\documentclass[twocolumn,superscriptaddress,showpacs,preprintnumbers,amsmath,amssymb,prl]{revtex4}
\usepackage[dvips]{graphicx}% Include figure files
\usepackage{dcolumn}% Align table columns on decimal point
\usepackage{bm}% bold math
\newcommand{\msr}{$\mu$SR}
\newcommand{\lco}{La$_2$CuO$_4$}

\newcommand{\lyco}{T'-La$_{1.9}$Y$_{0.1}$CuO$_4$}
\newcommand{\lycod}{T'-La$_{1.9}$Y$_{0.1}$CuO$_{4+\delta}$}
\begin{document}

\preprint{APS/123-QED}
\title{Bulk superconductivity in undoped \lyco\ probed by muon spin rotation}
% Force line breaks with \\

\author{K.~M. Kojima}
\affiliation{Muon Science Laboratory and Condensed Matter Research Center, Institute of Materials Structure Science, High Energy Accelerator Research Organization (KEK), Tsukuba, Ibaraki 305-0801, Japan}
\affiliation{Department of Materials Structure Science, The Graduate University for Advanced Studies, Tsukuba, Ibaraki 305-0801, Japan}
\author{Y. Krockenberger}
\affiliation{NTT Basic Research Laboratory, NTT Corporation, Atsugi, Kanagawa 243-0198, Japan}
\author{I. Yamauchi}
\affiliation{Muon Science Laboratory and Condensed Matter Research Center, Institute of Materials Structure Science, High Energy Accelerator Research Organization (KEK), Tsukuba, Ibaraki 305-0801, Japan}
\author{M. Miyazaki}
\affiliation{Muon Science Laboratory and Condensed Matter Research Center, Institute of Materials Structure Science, High Energy Accelerator Research Organization (KEK), Tsukuba, Ibaraki 305-0801, Japan}
\author{M. Hiraishi}
\affiliation{Muon Science Laboratory and Condensed Matter Research Center, Institute of Materials Structure Science, High Energy Accelerator Research Organization (KEK), Tsukuba, Ibaraki 305-0801, Japan}
\author{\\A. Koda}
\affiliation{Muon Science Laboratory and Condensed Matter Research Center, Institute of Materials Structure Science, High Energy Accelerator Research Organization (KEK), Tsukuba, Ibaraki 305-0801, Japan}
\affiliation{Department of Materials Structure Science, The Graduate University for Advanced Studies, Tsukuba, Ibaraki 305-0801, Japan}
\author{R. Kadono}\thanks{e-mail: ryosuke.kadono@kek.jp}
\affiliation{Muon Science Laboratory and Condensed Matter Research Center, Institute of Materials Structure Science, High Energy Accelerator Research Organization (KEK), Tsukuba, Ibaraki 305-0801, Japan}
\affiliation{Department of Materials Structure Science, The Graduate University for Advanced Studies, Tsukuba, Ibaraki 305-0801, Japan}
\author{R. Kumai}
\affiliation{Photon Factory and Condensed Matter Research Center, Institute of Materials Structure Science, High Energy Accelerator Research Organization (KEK), Tsukuba, Ibaraki 305-0801, Japan}
\affiliation{Department of Materials Structure Science, The Graduate University for Advanced Studies, Tsukuba, Ibaraki 305-0801, Japan}
\author{H. Yamamoto}
\affiliation{NTT Basic Research Laboratory, NTT Corporation, Atsugi, Kanagawa 243-0198, Japan}
%\author{H. Luetkens}
%\affiliation{Laboratory for Muon Spin Spectroscopy, Paul Scherrer Institute, CH-5232 Villigen PSI,  Switzerland}
%\author{A. Suter}
%\affiliation{Laboratory for Muon Spin Spectroscopy, Paul Scherrer Institute, CH-5232 Villigen PSI,  Switzerland}
\author{A. Ikeda}
\affiliation{Department of Applied Physics, Tokyo University of Agriculture and Technology, Koganei, Tokyo 184-8588, Japan}
\author{M. Naito}
\affiliation{Department of Applied Physics, Tokyo University of Agriculture and Technology, Koganei, Tokyo 184-8588, Japan}


\begin{abstract}

The Meissner effect has been directly demonstrated by depth-resolved muon spin rotation measurements in high-quality thin films of the T'-structured cuprate, \lyco, to confirm bulk superconductivity ($T_c\simeq21$ K) in its {\sl undoped} state. The gradual expelling of an external magnetic field is observed over a depth range of $\sim$100 nm in films with a thickness of 275(15) nm, from which the penetration depth is deduced to be 466(22) nm. Based on this result, we argue that the true ground state of the ``parent" compound of the $n$-type cuprates is not a Mott insulator but a strongly correlated metal with colossal sensitivity to apical oxygen impurities.
 
\end{abstract}

\maketitle

It seems now a canonical consensus in the physics of high-temperature superconductivity that the layered copper-oxide perovskites in their pristine state are Mott insulators.  
They are situated at the point of half-filling to the Cu $d_{x^2-y^2}$ band, where the band splits into upper and lower Hubbard bands (UHB/LHB) due to strong on-site Coulomb repulsion. 
Such band splitting freezes out charge degrees of freedom due to the fully occupied LHB (or, more precisely the charge-transferred band, CTB, consisting of oxygen $p$ orbitals above LHB \cite{Zaanen:85,Zhang:88}), and consequently it induces localized spins which interact magnetically via superexchange interaction that eventually leads to three-dimensional antiferromagnetic (AFM) order as a ground state.  Although such a picture based on the Hubbard model seems to capture most of the distinct properties of cuprates qualitatively, there are a number of subtleties that are subsided by this simplification.

One such subtle but not least essential feature lies in the structure of the CuO$_2$ planes which is slightly different between the hole-doped ($p$-type) and electron-doped ($n$-type) cuprates. The $p$-type compounds in the T-structure (close to that of K$_2$NiF$_4$, see Fig.~1a) has an apical oxygen as the Cu ions are six-fold coordinated, whereas the apical oxygen is absent in the $n$-type ones that crystalize into the so called T'-structure (Fig.~1b, space group $I4/mmm$) consisting of four-fold coordinated Cu ions. 
Practically, it has been known that the electronic properties of T'-structure cuprates are highly sensitive to off-stoichiometric oxygen impurities residing at the apical position \cite{Radaelli:94,Schultz:96}. The removal of apical oxygen to the level of $\simeq0.04$ per formula unit is crucial to obtain a bulk superconducting sample. While the AFM order observed in the underdoped region is believed to be of the Mott insulating phase realized in the {\sl stoichiometric} T'-structure cuprates, such attribution has been challenged by experiments on thin-film samples in which superconductivity is reported to occur even without carrier doping \cite{Matsumoto:09,Yamamoto:11,Tsukada:05,Krockenberger:12}. Consideration on the Madelung energy associated with the apical oxygen implies that there might be no excitation gap between UHB and CTB in the truly stoichiometric T'-structured cuprates \cite{Torrance:89,Ohta:91}, which has been partially supported by the recent theoretical investigation based on the local-density approximation combined with the dynamical mean field theory \cite{Das:09,Weber:10a,Weber:10b}.  These suggest the possibility that both AFM order and charge transfer gap are controlled by the apical oxygen impurities in $n$-type cuprates.

Here, we show by depth-resolved muon spin rotation (\msr) measurements that the superconductivity in the undoped thin films of \lyco\ (LYCO) is intrinsically of bulk nature, which is inferred from the gradual decrease of the measured local field with increasing depth in a manner predicted by the Meissner effect.  The result strongly suggests that the apical oxygen impurities in the T'-structure cuprates serve as a ``switch" of the electronic correlation, where superconductivity in the stoichiometric T'-structure emerges on the ground of a strongly correlated metal next to the Mott insulator.

High quality single-phase $c$-axis oriented thin-film samples of LYCO were grown by molecular beam epitaxy (MBE) on a $\langle110\rangle$ surface of YAlO$_3$ substrates with a thickness $2D=275(15)$ nm, where La$^{3+}$ was partially substituted with the isovalent Y$^{3+}$ to stabilize the T'-structure. This is needed since the T'-structure {\sl R}$_2$CuO$_4$ is unstable for {\sl R} = La due to its large ionic radius. The film surface was covered by a thin layer [$d_0=20(1)$ nm] of silver at the end of the growth process. The details of the film growth and subsequent oxygen reduction procedure are reported elsewhere \cite{Krockenberger:12,Naito:02}. The cation stoichiometry was adjusted to designed values by means of electron impact emission spectroscopy (EIES) during film growth, and the composition stoichiometry was verified by inductively coupled plasma analysis (ICP). As shown in Fig.~1c, the obtained thin films exhibit superconductivity with $T_c\simeq21$ K. The films were investigated by x-ray diffraction, transmission electron microscope (TEM), and synchrotron radiation (SR) x-ray measurements (at KEK-PF) to assess their quality, and they were confirmed to be single phase with less than 2-3\% of impurities (see Fig.~1d-1e). We note that the possibility of unintentional carrier doping is highly unlikely, as is inferred from various evidences including unchanged $a$-axis length and shrinkage of the $c$-axis during the reduction process  (see Supplementary Material \cite{Sup} for more detail). 

\begin{figure}[t]
	\centering
	\includegraphics[width=0.45\textwidth,clip]{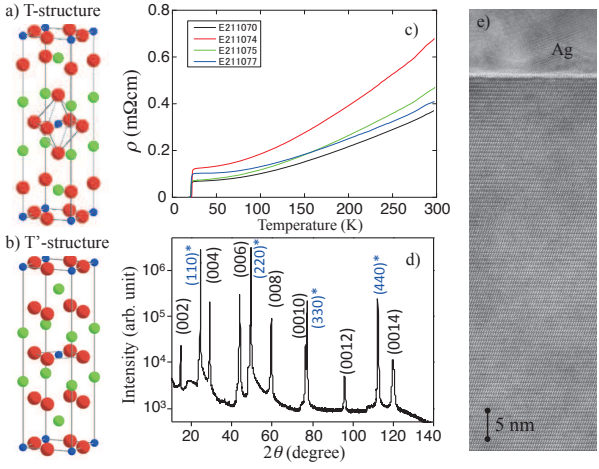}
	\caption{(Color online) Crystal structure of a) $p$-type and b) $n$-type {\sl R}$_2$CuO$_4$ (blue = Cu, red = O, green = {\sl R}). c) Resistivity of four thin-film samples of \lyco\ (LYCO) used for \msr\ measurement. d) x-ray (Cu-$K\alpha$) diffraction pattern of the thin-film sample obtained at the end of MBE growth, where peaks from LYCO film are observed besides those from YAlO$_3$ substrate (marked with asterisk). e) TEM image of the LYCO film near the film surface.} \label{fig1}
\end{figure}

 \msr\ measurements using low-energy muon, which are now routinely used for studies of superconductivity\cite{Jackson:00,Khasanov:04,Suter:04,Morenzoni:11}, have been performed at the Paul Scherrer Institut, Switzerland \cite{Morenzoni:04,Prokscha:08}. A muon beam with an energy ($E_\mu$) ranging from 4 to 23 keV was transported to sample position with an initial muon polarization $\vec{P}_\mu(0)$ perpendicular to the beam direction $\hat{z}$ [with $\vec{P}_\mu(0)\parallel \hat{x}$]. For each \msr\ measurement, a mosaic of four mostly identical $1.26\times1.26$ cm$^2$ thin films was used (see Fig.~1c for the resistivity of each film) to cover beam spot size of $2.5\times2.5$ cm$^2$.
During measurements for the Meissner effect, a weak magnetic field ($B_0\equiv\mu_0H\simeq7$ mT) was applied parallel to the sample surface [$\vec{H}\parallel \hat{y}$] after zero-field cooling down to $\sim$5 K.  Implanted muons probed the local transverse field (TF) which was modulated by the Meissner effect below $T_c$.  

For TF condition, $\vec{P}_\mu(t)$ rotates around the internal field $B=B(z)$ ($\parallel H$) so that the time evolution of an implanted muon polarization would be described by  $G_x(t)=\cos(\gamma_\mu B t+\phi)$, where $\gamma_\mu=2\pi\times 135.53$ MHz/T is the muon gyromagnetic ratio, and $\phi$ is the initial phase determined by the direction of $\vec{P}_\mu(0)$ relative to the $\hat{x}$ axis. 
Provided a model for $B(z)$, one can simulate the time evolution of the \msr\ spectra by a convolution of $B(z)$ and the muon depth distribution $n(z,E_\mu)$ calculated by a Monte Carlo code TRIM.SP \cite{trim:91},
\begin{equation} 
G_x(t) =  [\overline{n}(E_\mu)]^{-1}\int_0^\infty\cos[\gamma_\mu B(z) t+\phi]\cdot n(z,E_\mu)dz,\label{gxt}
\end{equation} 
and it is approximated by
\begin{equation} 
\overline{G}_x(t) = \simeq e^{-\frac{1}{2}\sigma^2t^2-\sigma_{\rm s}^2t^2}\cos(\gamma_\mu \overline{B}t+\phi),
\end{equation}
where $\overline{n}(E_\mu) = \int n(z,E_\mu)dz$, $\sigma$ and $\sigma_{\rm s}$ denote the depolarization due to the random local fields from nuclear magnetic moments and that due to the variation of $B(z)$, respectively, and $\overline{B}=\int B(z)n(z,E_\mu)dz/\overline{n}(E_\mu)$ is the average field. Although $n(z,E_\mu)$ is not symmetric with respect to $z$, the Gaussian approximation turns out to be quite satisfactory to reproduce the data.

We have found that there exists a signal component with an enhanced depolarization below $\sim$80 K, suggesting that mesoscopic, highly disordered magnetic domains develop with decreasing temperature. Moreover, it has been inferred from the preliminary analysis that the mean field $B_{\rm m}$ originating from the magnetism is considerably smaller than $B(z)$.  Assuming that superconductivity occurs in the non-magnetic domains (with a volume fraction $f_{\rm p}$) and that the depolarization in the magnetic domain is described by
\begin{equation}
G_{\rm m}^{\rm TF}(t)\simeq e^{-\frac{1}{2}\Lambda^2t^2}\cos[\gamma_\mu(\overline{B}+B_{\rm m})t+\phi],\label{gm}
\end{equation}
we analyzed the \msr\ spectra [i.e., time-dependent positron asymmetry, $A(t)\propto \hat{x}\cdot \vec{P}_\mu(t)$] by curve fits using a form
\begin{equation}
%A(t)=A_0[f_{\rm p}\overline{G}_x(t)+(1-f_{\rm p})e^{-\frac{1}{2}\Lambda^2t^2}],\label{asy}
A(t)=A_0[f_{\rm p}\overline{G}_x(t)+(1-f_{\rm p})G_{\rm m}^{\rm TF}(t)],\label{asy}
%G_{\rm m}(t)&=&\frac{1}{3}e^{-\Lambda_\parallel t}+\frac{2}{3}e^{-\frac{1}{2}\Lambda_\perp^2t^2}
\end{equation}
with $A_0$, $\overline{B}$, $B_{\rm m}$, $f_{\rm p}$, $\sigma_{\rm s}$, and $\Lambda$ as free parameters, where $A_0$ is the initial asymmetry and $\Lambda$ is the depolarization rate in the magnetic domains. ($\sigma$ was determined by the analysis of the ZF-\msr\ data: see below.)

\begin{figure}[t]
	\centering
	\includegraphics[width=0.45\textwidth,clip]{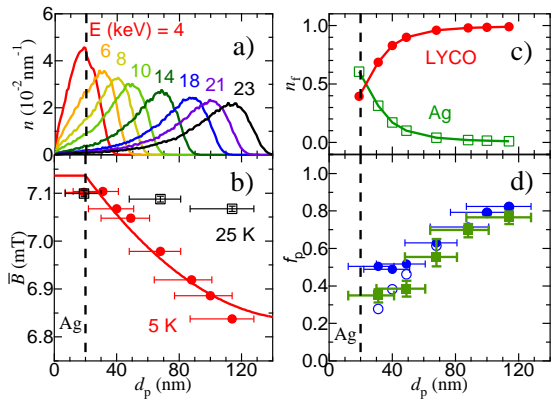}
	\caption{(Color online) a) Muon stopping profiles $n(z)$ in Ag-coated thin-film sample of \lyco\ (LYCO) for respective implantation energies ($E_\mu$) calculated by TRIM.SP. b) Mean local field $\overline{B}$ versus peak implantation depth $d_{\rm p}$ in LYCO probed by \msr\ under an external field $\mu_0H\simeq7$ mT, where filled circles are data at 5 K ($<T_c$) with solid curve obtained by fit using Eq.(\ref{bz}), and open squares are those at 25 K ($>T_c$). Horizontal bars correspond to FWHM of $n(z)$ at respective $E_\mu$. c) Simulated fractional yield $n_{\rm f}$ for muons stopping in LYCO (filled circles) and Ag layer (open squares). d) The non-magnetic volume fraction $f_{\rm p}$ determined by TF-  (filled circles) and ZF-\msr\ (filled squares) as a function of peak implantation depth $d_{\rm p}$ in LYCO. (Open circles denote $f_{\rm p}$ after considering $n_{\rm f(Ag)}$.)  The dashed line corresponds to the boundary between Ag coating and LYCO. } \label{fig2}
\end{figure}

The deduced mean local field $\overline{B}$ vs peak implantation depth $d_{\rm p}$ for $\mu_0H\simeq7$ mT is shown in Fig.~\ref{fig2}b, where one can observe a clear decrease of $\overline{B}$ at 5 K with increasing depth over a distance of $\sim$100 nm underneath the Ag layer. This magnetic field gradient is completely absent at 25 K ($>T_c$).  Since the magnetic penetration depth $\lambda$ ($=\lambda_{ab}$) is estimated to be of the same order as the film thickness ($2D$), the external field penetrates from both sides into the film where 
\begin{equation}
%B(z)=\frac{B_0}{1+e^{-D/\lambda}}[e^{-(z-d_0)/\lambda}+e^{(z-d_0-D)/\lambda}] \label{bz}
B(z)=B_0\frac{\cosh[(D-z+d_0)/\lambda]}{\cosh(D/\lambda)}\:\:\: ({\rm for}\:\:z\ge d_0) \label{bz}.
\end{equation}
A curve fit of the data in Fig.~\ref{fig2}b with this form yields $\lambda=466(22)$ nm and $B_0=7.14(2)$ mT.  Alternatively, one can perform curve fits using the time spectra given by Eqs.~(\ref{asy}), (\ref{gxt}) and (\ref{bz}) instead of $\overline{G}_x(t)$ (i.e., without resorting to $\overline{B}$ as an intermediate parameter), which yields $\lambda=465(5)$ nm. ($D$ and $d_0$ are fixed for both cases). These values are in excellent agreement with each other, and are in line with earlier measurements on T'-La$_{2-x}$Ce$_x$CuO$_4$ as they suggest a similar value of $\lambda$ for $x\rightarrow0$ \cite{Skinta:02}
It should be noted that the presence of magnetic domain near the surface might weaken the shielding current, so that the ``local $\lambda$" might be at variance from the above mentioned value. However, the observed depth profile of $\overline{B}$ is well reproduced by Eq.(\ref{bz}), suggesting that the model provides reasonable description of the Meissner effect  with $\lambda$ as a mean value averaged over $20\le d_{\rm p}\le120$ nm. 

%This corresponds to the in-plane penetration depth ($\lambda_{ab}$) typically found for the T-structure cuprates at the nominal carrier concentration of $\sim$0.1 per unit cell \cite{Uemura:89}. 

These analyses also yield the depth profile of the non-magnetic volume fraction $f_{\rm p}$  which is shown in  Fig.~\ref{fig2}c.  It exhibits a gradual increase with an increasing $d_{\rm p}$, being convex upward against $d_{\rm p}$. Considering the depth resolution limited by $n(z,E_\mu)$, we point out a remaining possibility that the actual depth profile of $f_{\rm p}$ may be sharper than that observed in Fig.~\ref{fig2}c near the surface, although the relatively small fraction of muon stopping in the LYCO film against that in Ag layer at low $E_\mu$ makes it difficult to assess the thickness of the magnetic domain in detail.
%Meanwhile, it is improbable that the profile comes from an antiferromagnetic monolayer at the surface, considering the small fraction of muons stopping within a monolayer LYCO ($\sim$1 nm).

%Furthermore, the depth dependence of $f_{\rm p}$ has a clear anti-correlation with $\overline{\lambda}$, indicating that the magnetic domains tend to suppress the superfluid density. 
%\begin{figure}[t]
%	\centering
%	\includegraphics[width=0.4\textwidth,clip]{fig3.eps}
%	\caption{(Color online) The ``local" penetration depth $\overline{\lambda}$ (open circles, left axis) determined by TF-\msr\ and non-magnetic volume fraction $f_{\rm p}$ (filled triangles, right axis) determined by ZF-\msr\ as a function of peak implantation depth $d_{\rm p}$ in LYCO. Horizontal errors correspond to FWHM of $n(z)$ at respective $E_\mu$. The region hatched by light-green corresponds to that of Ag coating. } \label{fig3}
%\end{figure}

\begin{figure}[t]
	\centering
	\includegraphics[width=0.45\textwidth,clip]{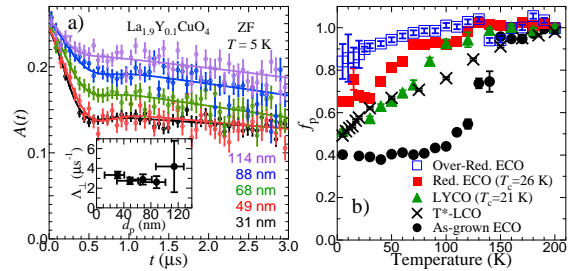}
	\caption{(Color online) a) ZF-\msr\ time spectra observed in LYCO at 5 K with various muon implantation depth ($d_{\rm p}$). Inset: Gaussian damping rate vs $d_{\rm p}$. b) Non-magnetic volume fraction $f_{\rm p}$ in MOD thin-film samples of T'-Eu$_2$CuO$_4$ (with 10 nm Au cap layer) determined by ZF-\msr\ after a variety of post-growth process ($E_\mu=4$ keV for Over-reduced and As-grown, 8 keV for Reduced).  Data of LYCO (measured at $E_\mu=6$ keV) and MBE-grown T$^*$-LCO ($E_\mu=12$ keV) are shown for comparison, where the latter is scaled with an offset ($\simeq0.5$) to reproduce $f_p$ in LYCO at the lowest temperature.} \label{fig3}
\end{figure}

The microscopic detail of the magnetic domain is inferred from depth-resolved ZF-\msr\ measurements, where the presence of a weak and highly disordered magnetism is identified near the sample surface. 
Figure \ref{fig3}a shows ZF-\msr\ time spectra at 5 K obtained with a variety of mean implantation depth $d_{\rm p}$, where the amplitude of a strongly damped oscillation exhibits a monotonic decrease with increasing $d_{\rm p}$. Curve fit analysis using 
%\begin{equation}
$A(t)=A_0[f_{\rm p}G_z(t)+(1-f_{\rm p})G_{\rm m}^{\rm ZF}(t)],$
%\end{equation}
with $G_z(t)\simeq e^{-\sigma^2t^2}$  and
$G_{\rm m}^{\rm ZF}(t)\simeq\frac{1}{3}e^{-\Lambda_\parallel t}+\frac{2}{3}e^{-\Lambda_\perp^2t^2}$ (i.e., an approximation of the Kubo-Toyabe function with $\Lambda_{\parallel,\perp}$ denoting the longitudinal and transverse depolarization rates due to the weak magnetism,  $\Lambda_{\perp}\simeq\Lambda$)
reproduces data, yielding parameter values that are consistent with those obtained from the TF data (details are found in the Supplementary Material \cite{Sup}).

We also performed additional \msr\ measurements on thin-film samples of (i) T'-Eu$_2$CuO$_4$ (ECO, $\sim$35$\pm$25 nm thick, with 10 nm Au coating) prepared by the metal-organic decomposition (MOD) technique \cite{Matsumoto:09} and (ii) T$^*$-structured La$_2$CuO$_4$ (T$^*$-LCO,  $\sim$120$\pm$5 nm thick) grown by MBE (see Supplement Material \cite{Sup}) to obtain clues for the weak magnetism in LYCO from their magnetic properties. As shown in Fig.~\ref{fig3}b,  $f_{\rm p}$ in the over-reduced sample (in which CuO$_2$ planes presumably accompany small amount of oxygen vacancies) is totally non-magnetic.    Meanwhile, the behavior of $f_{\rm p}$ in T$^*$-LCO is qualitatively similar to that in reduced ECO and LYCO, suggesting that the weak magnetism may come from the apical oxygen with a local structure closer to that of T$^*$-LCO. 

%while $f_{\rm p}$ for as-grown and reduced samples exhibits behavior quite similar to LYCO (near the surface), 
% Thus the oxygen defects in the CuO$_2$ plane, if at all, would not be the origin of the weak magnetism found in LYCO.(We note that all these samples were confirmed to retain their T'-structure by x-ray diffraction analysis.)

Considering the total thickness of LYCO film ($\simeq275$ nm), the lower bound for the net volume fraction of the magnetic phase is estimated to be $30/275$ - $50/275\simeq$11-18\%. Meanwhile, it is inferred from the SR x-ray analyses that the T-structure phase is absent with an upper bound of 2-3\%. This is also confirmed from the current \msr\ result, where the observed transition temperature for the weak magnetism is not higher than $\sim$80 K which is far below the N\'eel temperature ($\sim$300 K) of the T-\lco\ \cite{Reehius:06}. Thus one of the few remaining possibilities for the origin of the weak magnetism is  trace amount of oxygen at the apical position, which is plausible in view of the difficulty to control the stoichiometry with the relevant degree of precision. 

\begin{figure}[t]
	\centering
	\includegraphics[width=0.35\textwidth,clip]{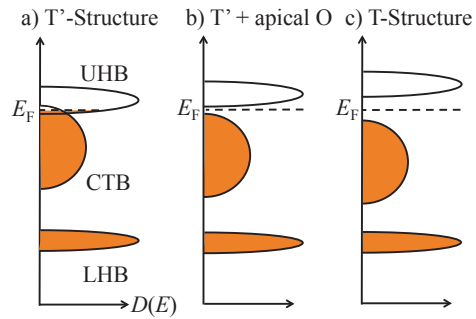}
	\caption{(Color online) A schematic diagram of band structures of La$_2$CuO$_{4+\delta}$ with a) stoichiometric T'-structure ($\delta=0$),  b) T'-structure with off-stoichiometric apical oxygen for $\delta\ge \delta_c$, and with c) T-structure, where UHB and LHB refer to the upper and lower Hubbard bands, and CTB to the charge transfer band. Both b) and c) take antiferromagnetic order as the ground state, while a) remains to be paramagnetic metal. } \label{fig4}
\end{figure}

Here, we make a conjecture that may explain the observed bulk superconductivity of \lycod\ including the origin of the near surface magnetism, where $\delta$ denotes the off-stoichiometric apical oxygen content.  Based on the bulk metallicity in the present samples, we presume that the true ground state of the ideal compound ($\delta=0$) is a metal in which UHB overlaps with CTB to yield the Fermi level situated in the overlapping region, as is depicted schematically in Fig.~\ref{fig4}a \cite{Adachi:13}.  %As mentioned earlier, this would be qualitatively justified by the consideration on the Madelung energy that predicts a lower UHB level for the T'-structure.  
The moderate carrier density inferred from \msr\ (as superfluid density) suggests that the overlap is rather marginal, making it sensitive to perturbation.  The situation resembles a semimetal (e.g., bismuth), where UHB and CTB serves as electron and hole bands, respectively.  

Now, introduction of an oxygen atom at the apical position might pull up the UHB level over a local domain of $m$ unit cells around it, so that the increase of $\delta$ above a critical content $\delta_c\simeq(2m)^{-1}$ would eventually lead to the departure of UHB from CTB with a small gap in-between to yield the bulk Mott insulating state with AFM order (Fig.~\ref{fig4}b).  Possibility that the apical oxygen leads to carrier doping is unlikely, as is inferred from insulating behavior of as-grown T'-films (i.e., $E_{\rm F}$ is unchanged). A recent report on a bulk crystal of T'-\lco\ grown by alkaline hydroxide flux technique suggests that $\delta\simeq0.018$ (estimated by x-ray Rietveld analysis) in its as-grown state is sufficient to induce AFM order over the entire volume of the sample \cite{Hord:10}. A detailed study on the magnetism of T'-PrLa$_{1-x}$Ce$_x$CuO$_4$ (PLCCO, which is obtained in bulk single crystal) as a function of oxygen deficiency revealed a remarkably sharp transition of the electronic state from bulk AFM order to a non-magnetic state at lower Ce content $x$ \cite{Kuroshima:03}, which is also consistent with the sensitivity to trace amount of apical oxygen. Conversely, these observations strongly suggest that $\delta_c\sim$1\% or less. Such ``colossal" sensitivity of the ground state to apical oxygen might be related to the bistable character suggested by recent theory \cite{Weber:10a,Weber:10b}. 

The observed weak magnetism may be then understood as due to staggered moments induced around the trace amount of apical oxygen impurities ($0<\delta<\delta_c$), which would mutually interact via RKKY interaction to undergo a spin-glass-like magnetic order as we observed in the near-surface region. A similar situation is suggested from a detailed study of oxygen reduction effects in PLCCO \cite{Adachi:13}. We point out that the onset temperature of weak magnetism in the superconducting samples of LYCO and ECO films is distinctly broad and lower than the N\'eel temperature of any T'-structured cuprates ($\ge 115$ K) at their as-grown state.  The observed tendency of segregation to the near-surface region may reflect the gradient of the apical oxygen concentration occurred during the post-growth process.

Finally, we stress that the bulk metallicity of stoichiometric T'-structured cuprates can be reconciled with the vast knowledge of $n$-type cuprates accumulated over the past decades \cite{Armitage:10}, as it would have been mostly concerned with the magnetism of apical oxygen-induced Mott insulating phase and associated superconductivity upon electron doping to it.  

 We would like to thank the PSI staff for their technical support during the $\mu$SR experiment. We also appreciate helpful discussion with Y. Koike, T. Adachi, M. Fujita,  K. Yamada, H. Kumigashira, Y. Murakami, A. Suter, and H. Luetkens. This work was partially supported by the Grant-in-Aid for Innovative Research Areas (23108004), MEXT, Japan.
%\end{acknowledgments}

\end{document}